\documentclass[11pt,a4paper]{article}
\usepackage{amsmath,amssymb}
\usepackage{epsfig,graphicx}
\begin{document}
\title{\bf Properties of the light scalar mesons face experimental data \thanks{Plenary session talk at
 QUARKS-2006,
Repino, St.Peterburg, May 19-25, 2006}}
\author{N.N. Achasov
\footnote{{\bf e-mail address}: achasov@math.nsc.ru}\\
 \small{\em
 Sobolev Institute for Mathematics}\\
\small{\em Academician Koptiug Prospekt, 4,
  Novosibirsk, 630090, Russia}}
  \date{\today}
 \maketitle
\begin{abstract}
The following topics are considered.
 \begin{enumerate}
\item Confinement, chiral dynamics, and light scalar
mesons
\item Chiral shielding of the $\sigma(600)$
\item The $\phi$ meson radiative decays about nature of light scalar
resonances
\item The $J/\psi$ decays about nature of light scalar
resonances
\item The $a_0(980)\to\gamma\gamma$ and
$f_0(980)\to\gamma\gamma$ decays about nature of light scalar
resonances
\item New round in $\gamma\gamma\to\pi^+\pi^-$, the Belle data
\item The  $a_0(980)-f_0(980)$ mixing:  theory and
experiment
\end{enumerate}
Arguments in favor of the four-quark model of the $a_0(980)$ and
$f_0(980)$ mesons are given.
\end{abstract}

\section{INTRODUCTION}

The scalar channel in the region up to 1 GeV became a stumbling
block of QCD. The point is that not only perturbation theory does
not work in these channels, but  sum rules as well  because there
are not solitary resonances in this region.  At the same time the
question on the nature of the light scalar mesons is major for
understanding the mechanism of the chiral symmetry realization,
arising from the confinement,  and hence for understanding the
confinement itself. In the talk are discussed the chiral shielding
of the $\sigma(600)$, $\kappa(800)$ mesons, a role of the
radiative $\phi$ decays, the heavy quarkonia decays, the
$\gamma\gamma$ collisions in decoding the nature of the light
scalar mesons and evidence in favor of the four-quark nature of
the light scalar mesons. New goal and objectives are considered
also.

To discuss actually the nature of the nonet of the light scalar
mesons: the  putative $f_0(600)$ (or $\sigma (600)$) and $\kappa
(700-900)$ mesons and the well-established $f_0(980)$ and
$a_0(980)$ mesons, one should explain not only their mass
spectrum, particularly the mass degeneracy of the $f_0(980)$ and
$a_0(980)$ states, but answer the next real challenges.
\begin{enumerate}
\item The copious $\phi\to\gamma f_0(980)$ decay
and especially the copious $\phi\to\gamma a_0(980)$ decay, which
looks as the decay plainly forbidden by the Okubo-Zweig-Iizuka
(OZI) rule in the quark-antiquark model of $a_0(980)=(u\bar u -
d\bar d)/\sqrt{2}$.
\item Absence of $J/\psi\to a_0(980)\rho$ and
$J/\psi\to f_0(980)\omega$ with copious $J/\psi\to a_2(1320)\rho$,
$J/\psi\to f_2(1270)\omega$ if $a_0(980)$ and $f_0(980)$ are $P$
wave states of $q\bar q$ like $a_2(1320)$ and $f_2(1270)$
respectively.
\item Absence of $J/\psi\to\gamma f_0(980)$ with copious
$J/\psi\to\gamma f_2(1270)$ and $J/\psi\to\gamma f_2^\prime
(1525)\phi$ if $f_0(980)$ is $P$ wave state of $q\bar q$ like
$f_2(1270)$ or $f_2^\prime(1525)$.
\item Suppression of $a_0(980)\to\gamma\gamma$ and $f_0(980)\to
\gamma\gamma$ with copious $a_2(1320)\to\gamma\gamma$,
$f_2(1270)\to\gamma\gamma$ if $a_0(980)$ and $f_0(980)$ are $P$
wave state of $q\bar q$ like $a_2(1320)$and $f_2(1270)$
respectively.
\end{enumerate}

 As Experiment suggests, Confinement forms colourless
observable hadronic fields and spontaneous breaking of chiral
symmetry with massless pseudoscalar fields. There are two possible
scenarios for QCD at low energy.\vspace*{0.15cm}

1. {\bf Non-linear \boldmath{$\sigma$}-model}\\[0.15cm]
$L=\frac{F_\pi^2}{2}Tr\left(\partial_\mu V(x)\partial^\mu
V^+(x)\right) +...,$ where $ V(x)=\exp \{2\imath\phi(x)/F_\pi\}.$
\vspace*{0.15cm}

2. {\bf Linear \boldmath{$\sigma$}-model}\\[0.15cm]
$L=\frac{1}{2}Tr\left(\partial_\mu \mbox{V}(x)\partial^\mu
\mbox{V}^+(x)\right)- W\left(\mbox{V}(x)\mbox{V}^+(x)\right),$
where $\mbox{V}(x)=(\sigma(x)+\imath\pi(x)).$ \vspace*{0.15cm}

The experimental nonet of the light scalar mesons [ the putative
$f_0(600)$ (or $\sigma (600)$) and $\kappa (700-900)$ mesons and
the well-established $f_0(980)$ and $a_0(980)$ mesons] as if
suggests the $U_L(3)\times U_R(3)$ linear $\sigma$ model. Hunting
the light $\sigma$ and $\kappa$ mesons had begun in the sixties
already and a preliminary information on the light scalar mesons
in Particle Data Group (PDG) Reviews had appeared at that time.
But long-standing unsuccessful attempts to prove their existence
in a conclusive way entailed general disappointment and an
information on these states disappeared from PDG Reviews. One of
principal reasons against the $\sigma$ and $\kappa$ mesons was the
fact that both $\pi\pi$ and $\pi K$ scattering phase shifts do not
pass over $90^0$ at putative resonance masses.

\section{CHIRAL SHIELDING OF THE \boldmath{$\sigma(600)$} \cite{AS1,ADSK,AK}}

Situation changes when we showed that in the linear $\sigma$ model
\cite{GML},\\[0.15cm] $L=\left(1/2\right)\left
[(\partial_\mu\sigma)^2+(\partial_\mu\overrightarrow{\pi})^2\right]
+(\mu^2/2)[(\sigma)^2+(\overrightarrow{\pi})^2]-(\lambda/4)
[(\sigma)^2+(\overrightarrow{\pi})^2]^2,$\\[0.15cm] there is a
negative background phase which hides the $\sigma$ meson
\cite{AS1}. It has been made clear that shielding wide lightest
scalar mesons in chiral dynamics is very natural. This idea was
picked up and triggered new wave of theoretical and experimental
searches for the $\sigma$ and $\kappa$ mesons. According the
simplest Dyson equation for the $\pi\pi$ scattering amplitude with
real intermediate $\pi\pi$ states only, see Fig. 1,

\begin{figure}[h]
\includegraphics[width=34pc]{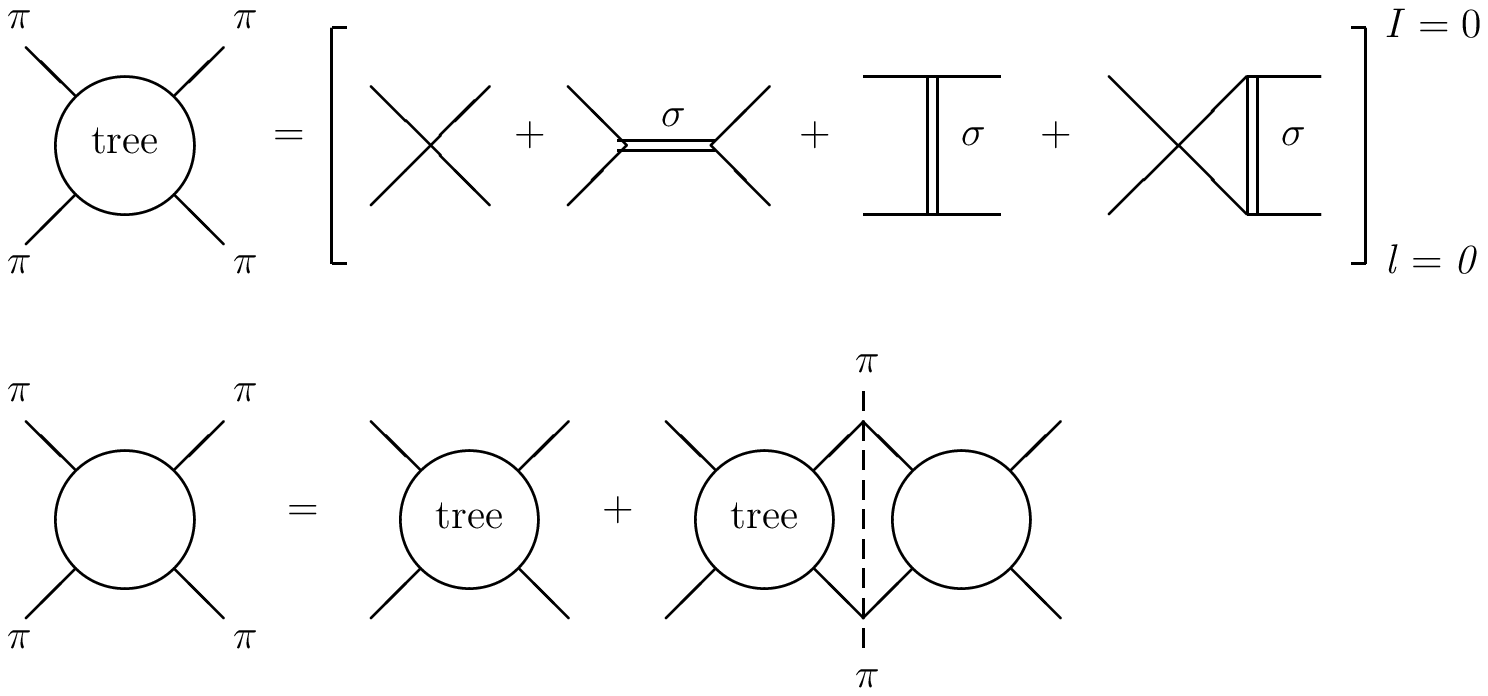}
\caption{\small The graphical representation of the $S$ wave $I=0$
$\pi\pi$ scattering amplitude.}
\end{figure}

$$ T^0_0=\frac{T_0^{0(tree)}}{1-\imath\rho_{\pi\pi}T_0^{0(tree)}}=
\frac{e^{2\imath\delta^0_0}-1}{2\imath \rho_{\pi\pi}}
=\frac{e^{2\imath\delta_B^{\pi\pi}}-1}{2\imath\rho_{\pi\pi}}+
e^{2\imath\delta_B^{\pi\pi}}T_{res}\,,$$\\
 $$
\delta^0_0=\delta_B^{\pi\pi}+\delta_{res}\,,\ \
\rho_{\pi\pi}=\sqrt{1-4m_\pi^2/s}\,,$$\\
 $$
T_{res}=\frac{1}{\rho_{\pi\pi}}\left [
\frac{\sqrt{s}\Gamma_{res}(s)}{M^2_{res} - s +
\Re(\Pi_{res}(M^2_{res}))- \Pi_{res}(s)}\right ]
=\frac{e^{2\imath\delta_{res}}-1}{2\imath\rho_{\pi\pi}}\,,$$\\ $$
T_B=\frac{\lambda(s)}{1-\imath\rho_{\pi\pi}\lambda(s)}
=\frac{e^{2\imath\delta_B^{\pi\pi}}-1}{2\imath \rho_{\pi\pi}}\,, \
\  \lambda(s)= \frac{m_\pi^2-m_\sigma^2}{32\pi F^2_\pi}\left
[5-2\frac{m_\sigma^2-m_\pi^2}{s-4m_\pi^2}\ln\left
(1+\frac{s-4m^2_\pi}{m_\sigma^2}\right )\right ]\,,$$\\
 $$
\Re(\Pi_{res}(s))=-3\frac{g_{res}^2(s)}{32\pi}\lambda(s)
\rho_{\pi\pi}^2\,,\ \
\Im(\Pi_{res}(s))=\sqrt{s}\Gamma_{res}(s)=\frac{3}{2}\frac{g_{res}^2(s)}
{16\pi}\rho_{\pi\pi}\,,$$\\
  $$ M^2_{res}= m_\sigma^2 -
\Re(\Pi_{res}(M^2_{res}))\,,\ \
g_{res}(s)=\frac{g_{\sigma\pi^+\pi^-}}{\bigr
|1-\imath\rho_{\pi\pi}\lambda(s)\bigl |}\,, $$  where $s=m^2$ and
$m$ is the invariant mass of the $\pi\pi$ system. These simple
formulae show that the resonance contribution is strongly modified
by the chiral background amplitude.

\begin{figure}[h]
\begin{center}
\begin{tabular}{ccc}
\hspace*{-12pt}\includegraphics[width=13pc]{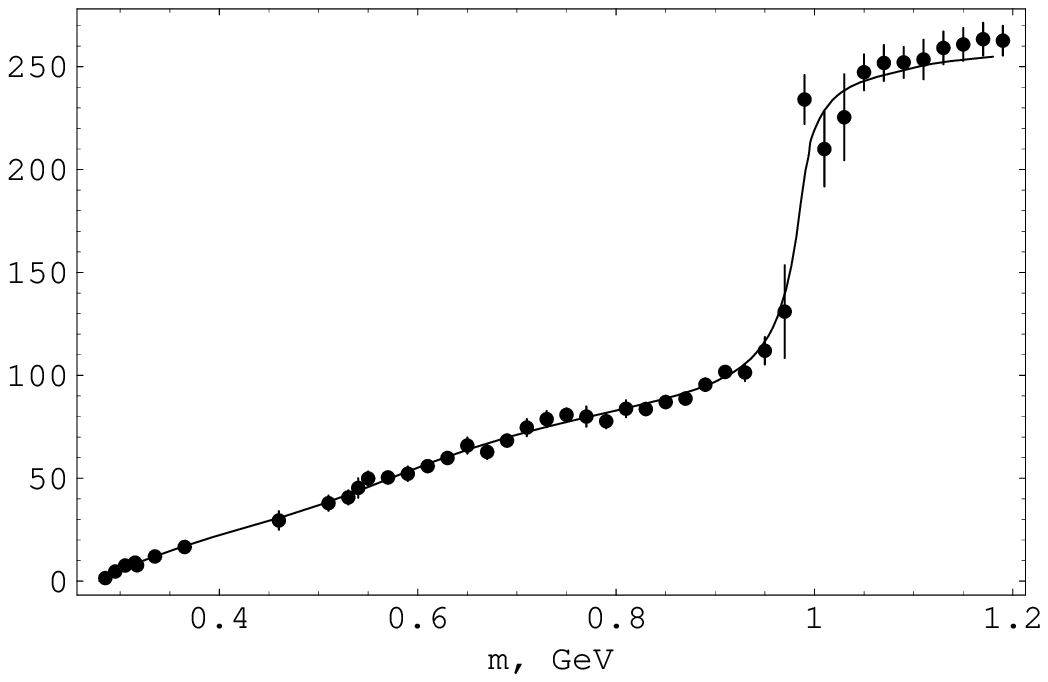}&
\hspace*{-12pt}\includegraphics[width=13pc]{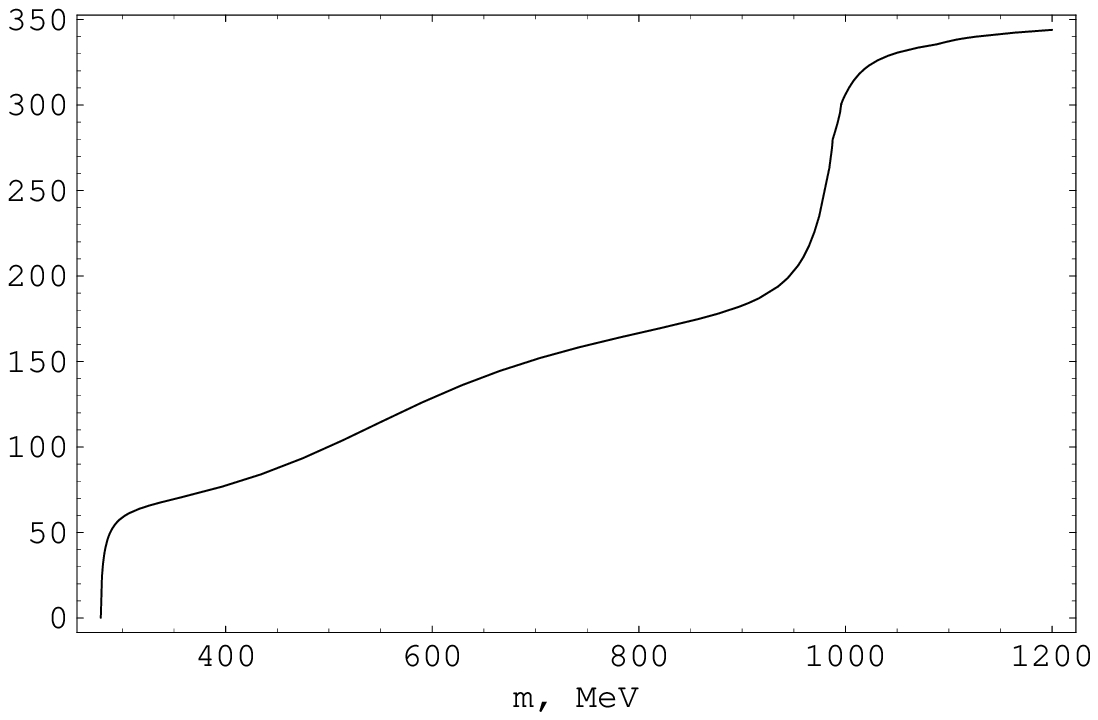}&
\hspace*{-12pt}\includegraphics[width=13pc]{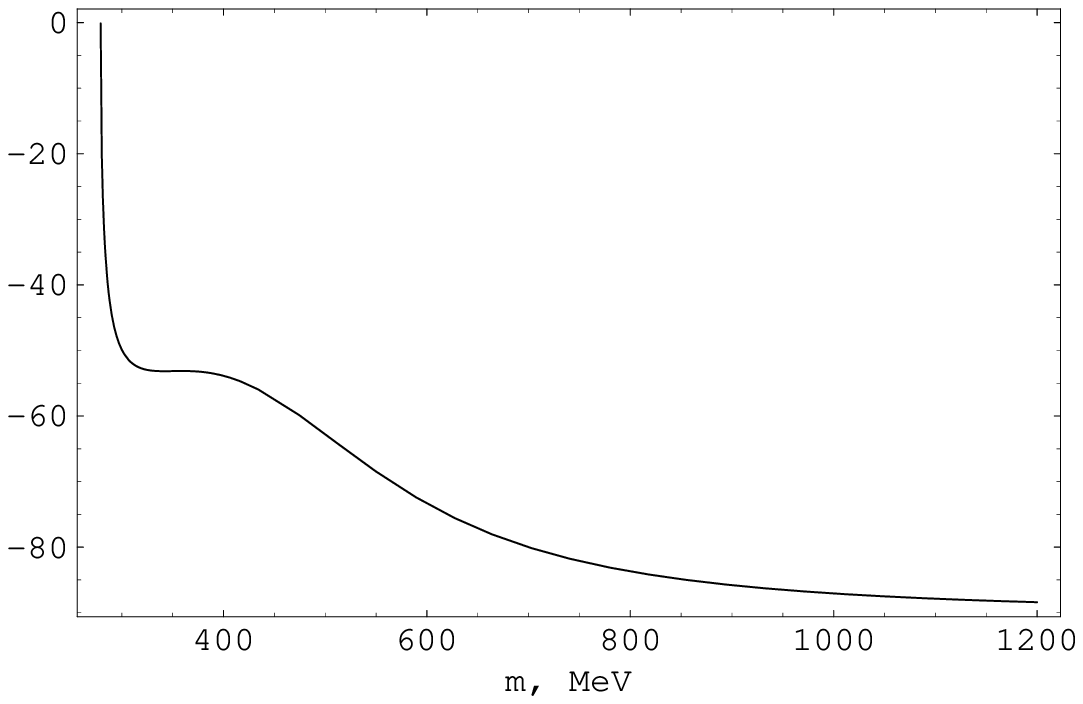}\\
(a)&(b)&(c)
\end{tabular}
\end{center}
\caption{\small\hspace*{12pt} (a) The $S$ wave $I=0$ $\pi\pi$
scattering phase shift $\delta_0^0$. (b) The resonance phase shift
$\delta_{res}$. (c) The background phase shift
$\delta_B^{\pi\pi}$.}
 \label{pipi}
\end{figure}

In theory the principal problem is impossibility to use the linear
$\sigma$ model in the tree level approximation inserting widths
into $\sigma$ meson propagators because such an approach breaks
both unitarity and Adler self-consistency conditions. Strictly
speaking, the comparison with the experiment requires the
non-perturbative calculation of the process amplitudes.
Nevertheless, now there are the possibilities to estimate odds of
the $U_L(3)\times U_R(3)$ linear $\sigma$ model to underlie
physics of light scalar mesons in phenomenology. Really, even now
there is a huge body of information about the $S$ waves of
different two-particle pseudoscalar states. As for theory, we know
quite a lot about the scenario under discussion: the nine scalar
mesons, the putative chiral shielding of the $\sigma (600)$ and
$\kappa(700-900)$ mesons, the unitarity, analiticity and Adler
self-consistency conditions. In addition, there is the light
scalar meson treatment motivated by field theory. The foundations
of this approach were formulated in our papers \cite{ADSK}. In
particular, in this approach  were introduced propagators of
scalar mesons, satisfying the K\"allen-Lehmann representation.
Recently \cite{AK}, the comprehensive examination of the chiral
shielding of the $\sigma (600)$ has been performed with a
simultaneous analysis of the modern data on the
$\phi\to\gamma\pi^0\pi^0$ decay and the classical $\pi\pi$
scattering data. Figs. \ref{pipi} (a), \ref{pipi} (b), and
\ref{pipi} (c) show an example of the fit to the data on the $S$
wave $I=0$ $\pi\pi$ scattering phase shift
$\delta_0^0=\delta_B^{\pi\pi}+ \delta_{res}$, the resonance
($\delta_{res}$) and  background ($\delta_B^{\pi\pi}$) components
of $\delta_0^0$, respectively (all the phases in degrees).

An example of the fit to the $\phi\to\gamma\pi^0\pi^0$ data in
this case is shown in Fig. 6.

\section{FOUR-QUARK MODEL}

The nontrivial nature of the well-established light scalar
resonances $f_0(980)$ and $a_0(980)$ is no longer denied
practically anybody. In particular, there exist numerous evidences
in favour of the $q^2\bar q^2$ structure of these states
\cite{A1,A2}. As for the nonet as a whole, even a look at PDG
Review gives an idea of the four-quark structure of the light
scalar meson nonet, $\sigma (600)$, $\kappa (700-900)$,
$f_0(980)$, and $a_0(980)$, inverted in comparison with the
classical $P$ wave $q\bar q$ tensor meson nonet, $f_2(1270)$,
$a_2(1320)$, $K_2^\ast(1420)$, $f_2^\prime (1525)$. \footnote{To
be on the safe side, notice that the linear $\sigma$ model does
not contradict to non-$q\bar q$ nature of the low lying scalars
because Quantum Fields can contain different virtual particles in
different regions of virtuality.}

Really, while the scalar nonet cannot be treated as the $P$ wave
{$q\bar q$ nonet in the naive quark model, it can be easy
understood as the $q^2\bar q^2$ nonet, where $\sigma (600)$ has no
strange quarks, $\kappa (700-900)$ has the $s$ quark, $f_0(980)$
and $a_0(980)$ have the $s\bar s$ pair \cite{J,BFSS}. The scalar
mesons $a_0(980)$ and $f_0(980)$, discovered more than thirty
years ago, became the hard problem for the naive $q\bar q$ model
from the outset. Really, on the one hand the almost exact
degeneration of the masses of the isovector $a_0(980)$ and
isoscalar $f_0(980)$ states revealed seemingly the structure
similar to the structure of the vector $\rho$ and $\omega$ or
tensor $a_2(1320)$ and $f_2(1270)$ mesons, but on the other hand
the strong coupling of $f_0(980)$ with the $K\bar K$ channel as if
suggested a considerable part of the strange pair $s\bar s$ in the
wave function of $f_0(980)$. In 1977 R.L. Jaffe \cite{J} noted
that in the MIT bag model, which incorporates confinement
phenomenologically, there are light four-quark scalar states. He
suggested also that $a_0(980)$ and $f_0(980)$ might be these
states with symbolic structures: $a^0_0(980)=(us\bar u\bar
s-ds\bar d\bar s)/\sqrt{2}$ and $f_0(980)=(us\bar u\bar s + ds\bar
d\bar s)/\sqrt{2}$. From that time $a_0(980)$ and $f_0(980)$
resonances came into beloved children of the light quark
spectroscopy.

\section{RADIATIVE DECAYS OF \boldmath{$\phi$} MESON ABOUT NATURE OF LIGHT
SCALAR RESONANCES \cite{AI,A2,AG}}

Ten years later we showed \cite{AI} that the study of the
radiative decays $\phi\to\gamma a_0\to\gamma\pi\eta$ and
$\phi\to\gamma f_0\to \gamma\pi\pi$ can shed light on the problem
of $a_0(980)$ and $f_0(980)$ mesons. Over the next ten years
before experiments (1998) the question was considered from
different points of view. Now these decays have been studied not
only theoretically, but  experimentally as well by energies of the
SND,  CMD-2, and KLOE.

\begin{eqnarray}
&& BR(\phi\to\gamma\pi^0\eta)= (0.83\pm0.05)\cdot
10^{-4}\nonumber\\[9pt]
 && BR(\phi\to\gamma\pi^0\pi^0)=
(1.09\pm0.06)\cdot10^{-4}\nonumber
\end{eqnarray}}

  Note that $a_0(980)$  is produced in the radiative
$\phi$ meson decay as intensively as $\eta '(958)$  containing
$\approx 66\% $ of $s\bar s$, responsible for $\phi\approx s\bar
s\to\gamma s\bar s\to\gamma \eta '(958)$. It is a clear
qualitative argument for the presence of the $s\bar s$ pair in the
isovector $a_0(980)$ state, i.e., for its four-quark nature.

\begin{figure}[h]
\includegraphics[width=36pc]{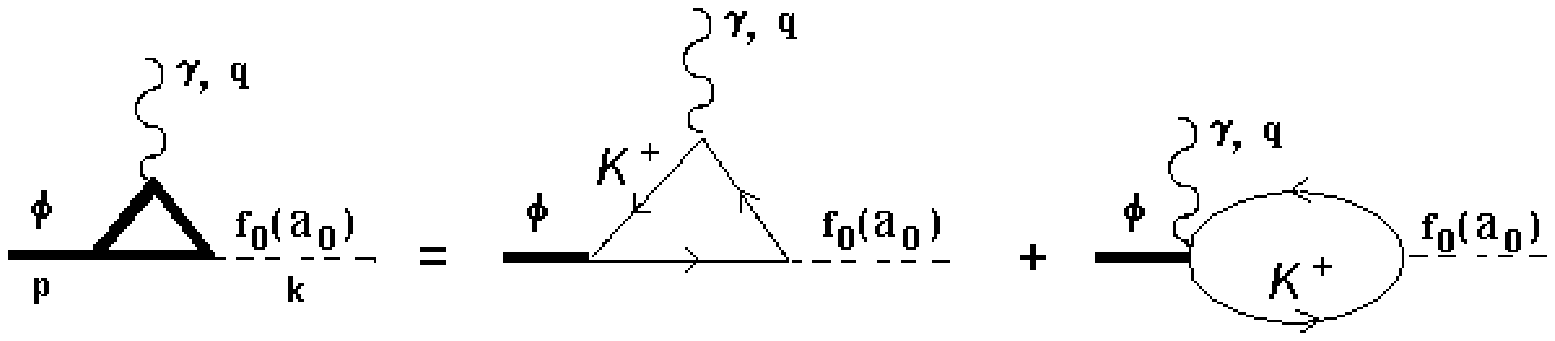}
\caption{\small Diagrams of the $K^+K^-$ loop model.}
 \label{model}
\end{figure}

When basing the experimental investigations, we suggested one-loop
model $\phi\to K^+K^-\to\gamma a_0(980)$ (or $f_0(980))$
\cite{AI}, see Fig. \ref{model}.

\begin{figure}[h]
\begin{center}
\begin{tabular}{ccc}
\hspace*{-14pt}\includegraphics[width=13pc]{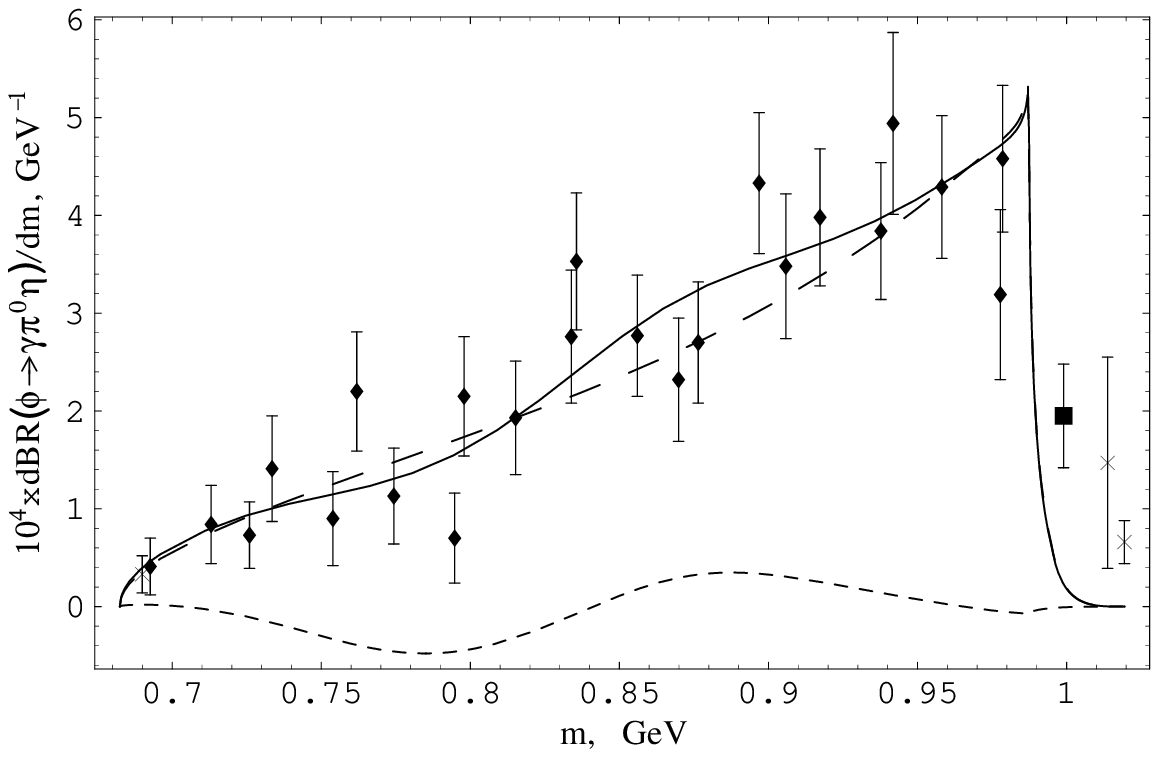}&
\hspace*{-12pt}\includegraphics[width=13pc]{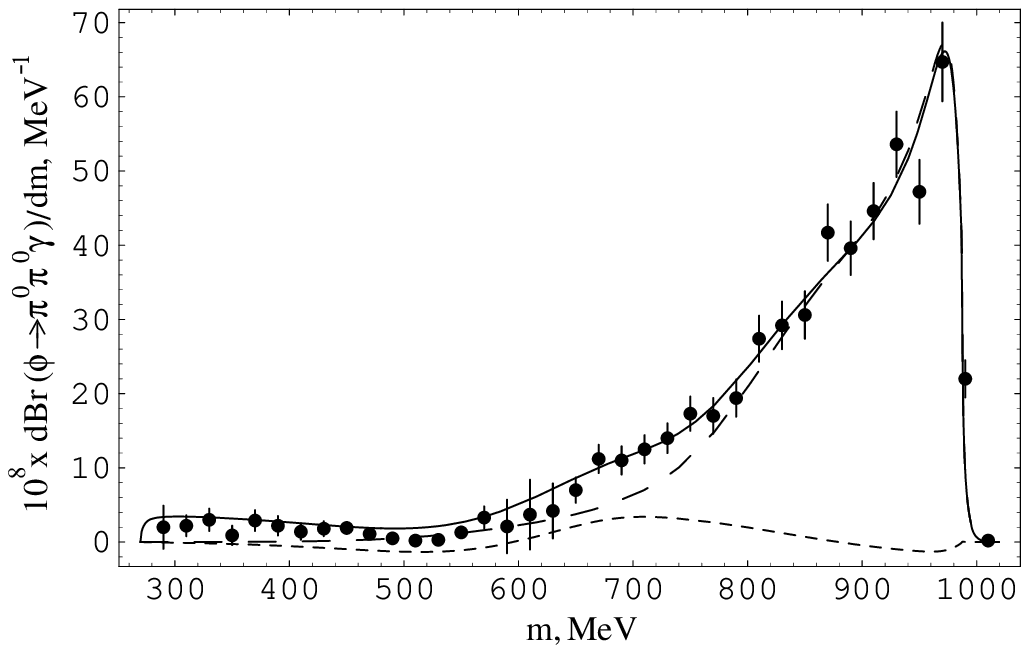}&
\hspace*{-12pt}\includegraphics[width=13pc]{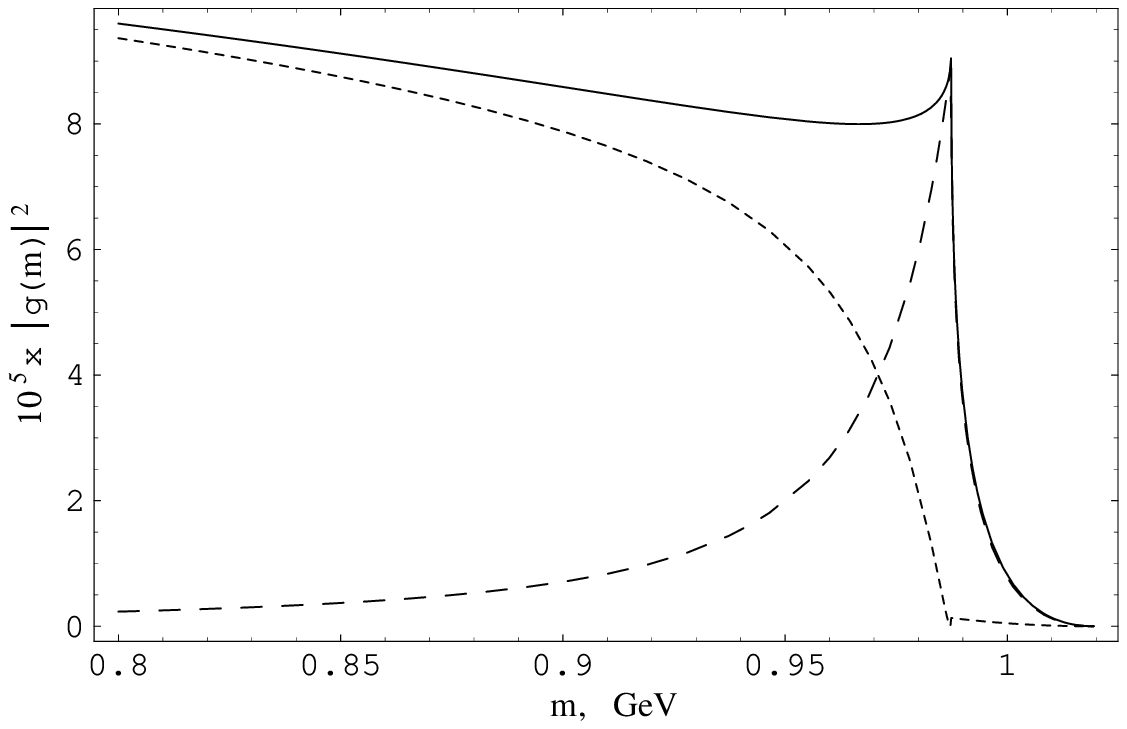}\\ (a)&(b)&(c)
\end{tabular}
\end{center}
\caption{\small (a) The fit to the KLOE data for the $\pi^0\eta$
mss spectrum in the $\phi\to\gamma\pi^0\eta$ decay. (b) The fit to
the KLOE data for the $\pi^0\pi^0$ mss spectrum in the
$\phi\to\gamma\pi^0\pi^0$ decay. (c) The universal in the $K^+K^-$
loop model function $|g(m)|^2=|g_R(m)/g_{RK^+K^-}|^2$ is shown by
the solid curve. The contribution of the imaginary (real) part is
shown by dashed (dotted) curve.}
 \label{phigammaR}
\end{figure}

 This model is used in the data treatment and is
ratified by experiment. Below we argue on gauge invariance grounds
that the present data give the conclusive arguments in favor of
the $K^+K^-$ loop transition as the principal mechanism of
$a_0(980)$ and $f_0(980)$ meson production in the $\phi$ radiative
decays \cite{A2,AG}. The data are described in the model
$\phi\to(\gamma a_0+\pi^0\rho)\to\gamma\pi^0\eta$ and $\phi\to
[\gamma (f_0+\sigma)+\pi^0\rho]\to\gamma\pi^0\pi^0$
\cite{AK,AG,AG1}. The resulting fits to the KLOE data are
presented in Figs. \ref{phigammaR} (a) and \ref{phigammaR} (b)
\footnote{The last KLOE investigation \cite{kloe} with the
statistics corresponding to 450 pb${^-1}$ supports our \cite{AK}
analysis.}.

 To
describe the experimental spectra $$S_R(m)\equiv dB(\phi\to\gamma
R\to\gamma ab\,,\, m)/dm$$
\begin{eqnarray}&& =\frac{2m^2\Gamma(\phi\to\gamma
R\,,\,m)\Gamma(R\to ab\,,\,m)}{\pi\Gamma_\phi|D_R(m)|^2} =
\frac{4|g_R(m)|^2\omega (m) p_{ab}(m)}{\Gamma_\phi\,
3(4\pi)^3m_{\phi}^2}\left |\frac{g_{Rab}}{D_R(m)}\right
|^2\nonumber
\end{eqnarray} (where $1/D_R(m)$ and  $g_{Rab}$ are the propagator
and coupling constants of $R=a_0,f_0$; $ab=\pi^0\eta$,
$\pi^0\pi^0$) the function $|g_R(m)|^2$ should be smooth, almost
constant, in the range $m\leq 0.99$ GeV. But the problem issues
from gauge invariance which requires that $$ A [\phi(p)\to\gamma
(k) R(q)]= G_R(m)[p_\mu e_\nu(\phi) - p_\nu e_\mu(\phi)][k_\mu
e_\nu(\gamma) - k_\nu e_\mu(\gamma)].$$
 Consequently, the function
$$g_R(m)= - 2(pk)G_R(m) = - 2\omega (m) m_\phi G_R(m)$$
 is
proportional to the photon energy
$\omega(m)=(m_{\phi}^2-m^2)/2m_{\phi}$ (at least!) in the soft
photon region. Stopping the function $(\omega (m))^2$ at $\omega
(990\,\mbox{MeV})=29$ MeV with the help of the form-factor
$1/\left [1+(R\omega (m))^2\right ]$ requires $R\approx 100$
GeV$^{-1}$. It seems to be incredible to explain such a huge
radius in hadron physics. Based on rather great $R\approx 10$
GeV$^{-1}$, one can obtain an effective maximum of the mass
spectrum only near 900 MeV. To exemplify this trouble let us
consider the contribution of the isolated $R$ resonance:
$g_R(m)=-2\omega (m) m_\phi G_R\left (m_R\right)$. Let also the
mass and the width of the $R$ resonance equal 980 MeV and 60 MeV,
then $S_R(920\,\mbox{MeV}):
S_R(950\,\mbox{MeV}):S_R(970\,\mbox{MeV}):S_R(980\,\mbox{MeV})
=3:2.7:1.8:1$. So stopping the $g_R(m)$ function is the crucial
point in understanding  the mechanism  of the production of
$a_0(980)$ and $f_0(980)$ resonances in the $\phi$ radiative
decays. The $K^+K^-$-loop model $\phi\to K^+K^-\to\gamma R$ solves
this problem in the elegant way: fine threshold phenomenon is
discovered, see Fig. \ref{phigammaR} (c). So, the mechanism of
production of $a_0(980)$ and $f_0(980)$ mesons in the $\phi$
radiative decays is established
 at a physical level of proof, see
Refs. \cite{A2,AG} for details.

 Both real and imaginary parts of the
$\phi\to\gamma R$
 amplitude are caused by the $K^+K^-$ intermediate state, see Fig. \ref{phigammaR} (c). The
imaginary part is caused by the real  $K^+K^-$ intermediate state
while the real part is caused by the virtual compact  $K^+K^-$
intermediate state, i.e., we are dealing here with  the four-quark
transition.

 Needless to say, radiative four-quark
transitions can happen between two  $q\bar q$ states as well as
between  $q\bar q$ and  $q^2\bar q^2$ states but their intensities
depend strongly on a type of the transitions.

 A radiative four-quark transition between two
 $q\bar q$ states requires creation and annihilation of an
additional  $q\bar q$ pair, i.e., such a transition is forbidden
according to the  Okubo-Zweig-Iizuka (OZI) rule, while a radiative
four-quark transition between  $q\bar q$ and  $q^2\bar q^2$ states
requires only creation of an additional  $q\bar q$ pair, i.e.,
such a transition is allowed according to the  OZI rule.

 The
four-quark transition  constrains the large $N_C$ expansion of the
$\phi\to\gamma a_0(980)$ and $\phi\to\gamma f_0(980)$ amplitudes
and gives the new strong (if not crucial) evidences in favor of
the compact four-quark nature of $a_0(980)$ and $f_0(980)$ mesons:
$a_0^0=(u\bar ss\bar u-d\bar ss\bar d)/\sqrt{2}$,  $f_0^0=(u\bar
ss\bar u+d\bar ss\bar d)/\sqrt{2}$, similar (but, generally
speaking, not identical) the MIT-bag states \cite{A2}.

\section{THE \boldmath{$J/\psi$} DECAYS AND THE
$a_0(980)\to\gamma\gamma$,  $f_0(980)\to\gamma\gamma$ DECAYS ABOUT
NATURE OF LIGHT SCALAR RESONANCES \cite{A1}}

{\bf The \boldmath $a_0(980)$ in $J/\psi$ decays.} The following
data is of very interest for our purposes: $B(J/\psi\to
a_0(980)\rho) < 4.4\cdot 10^{-4}$ and $B(J/\psi\to a_2(1320)\rho)=
(109\pm 22)\cdot 10^{-4}$. The suppression $B(J/\psi\to
a_0(980)\rho)/B(J/\psi\to a_2(1320)\rho)<0.04\pm0.008$ seems
strange, if one considers the $a_2(1320)$ and $a_0(980)$ states as
the tensor and scalar isovector states from the same $P$-wave
$q\bar q$ multiplet. While the four-quark nature of the $a_0(980)$
meson is not contrary to the suppression under discussion. So, the
improvement of the upper limit for $B(J/\psi\to a_0(980)\rho)$ and
the search for the $J/\psi\to a_0(980)\rho$ decays  are the urgent
purposes in the study of the $J/\psi$ decays!

Recall that twenty years ago the four-quark nature of $a_0(980)$
was supported by suppression of $a_0(980)\to\gamma\gamma$ as was
predicted in our work based on the $q^2\bar q^2$ model,
$\Gamma(a_0(980)\to\gamma\gamma)\sim 0.27\,\mbox{keV}$. Experiment
gives $\Gamma (a_0\to\gamma\gamma)=(0.19\pm 0.07
^{+0.1}_{-0.07})/B(a_0\to\pi\eta)$ keV, Crystal Ball, and $\Gamma
(a_0\to\gamma\gamma)=(0.28\pm 0.04\pm 0.1)/B(a_0\to\pi\eta)$ keV,
JADE. When in the $q\bar q$ model it was anticipated
$\Gamma(a_0\to\gamma\gamma)=(1.5 - 5.9)\Gamma
(a_2\to\gamma\gamma)= (1.5 - 5.9)\cdot(1.04\pm 0.09)$ keV.

{\bf The \boldmath $f_0(980)$ in $J/\psi$ decays.} The hypothesis
that the $f_0(980)$ meson is the lowest two-quark $P$ wave scalar
state with the quark structure $f_0(980)=(u\bar u+d\bar
d)/\sqrt{2}$ contradicts the following facts. 1) The strong
coupling with the $K\bar K$-channel,
$1<|g_{f_0K^+K^-}/g_{f_0\pi^+\pi^-}|^2<10$, for the prediction
$|g_{f_0K^+K^-}/g_{f_0\pi^+\pi^-}|^2=\lambda/4\simeq 1/8$. 2) The
weak coupling with gluons, $B(J/\psi\to\gamma
f_0(980)\to\gamma\pi\pi)< 1.4\cdot10^{-5}$, opposite the expected
one $B(J/\psi\to\gamma f_0(980))\approx B(J/\psi\to\gamma
f_2(1270))/4=(3.45\pm 0.35)\cdot 10^{-4}$. 3) The weak coupling
with photons, predicted in our work for the $q^2\bar q^2$ model,
$\Gamma(f_0(980)\to\gamma\gamma)\sim 0.27\,\mbox{keV}$, and
supported by experiment, $\Gamma (f_0\to\gamma\gamma)=(0.31\pm
0.14\pm 0.09)$ keV, Crystal Ball, and $\Gamma
(f_0\to\gamma\gamma)=(0.24\pm 0.06\pm 0.15)$ keV, MARK II. When in
the $q\bar q$ model it was anticipated
$\Gamma(f_0\to\gamma\gamma)=(1.7 - 5.5)\Gamma
(f_2\to\gamma\gamma)= (1.7 - 5.5)\cdot(2.8\pm 0.4)$ keV. 4) As is
the case with $a_0(980)$ the suppression $B(J/\psi\to
f_0(980)\omega) /B(J/\psi\to f_2(1270)\omega)= 0.033\pm 0.013$
looks strange in the model under consideration. We should like to
emphasize that from our point of view the DM2 Collaboration did
not observed the $J/\psi\to f_0(980)\omega$ decay and should give
a upper limit only. So, the search for the $J/\psi\to
f_0(980)\omega$ decay is the urgent purpose in the study of
the$J/\psi$ decays! The existence of the $J/\psi\to f_0(980)\phi$
decay of greater intensity than the $J/\psi\to f_0(980)\omega$
decay shuts down the $f_0(980)=(u\bar u+d\bar d)/\sqrt{2}$ model.
In the case under discussion the $J/\psi\to f_0(980)\phi$ decay
should be strongly suppressed in comparison with the $J/\psi\to
f_0(980)\omega$ decay by the OZI rule.

Can one consider the $f_0(980)$ meson as the near $s\bar s$-state?
It is impossible without a gluon component. Really, it is
anticipated for the scalar $s\bar s$-state from the lowest P-wave
multiplet that $B(J/\psi\to\gamma f_0(980))\approx
B(J/\psi\to\gamma f_2^\prime(1525))/4=(1.175
^{+0.175}_{-0.125})\cdot 10^{-4}$ opposite experiment $< 1.4\cdot
10^{-5}$, which requires properly that the $f_0(980)$-meson to be
the 8-th component of the $SU_f(3)$ oktet $f_0(980)=(u\bar u+d\bar
d-2s\bar s)/\sqrt{6}$. But this structure gives $B(J/\psi\to
f_0(980)\phi)=(2\lambda\approx 1)\cdot B(J/\psi\to f_0(980)
\omega)$ which is on the verge of conflict with experiment. Here
$\lambda$ takes into account the strange sea suppression. The
$SU_f(3)$ oktet case  contradicts also the strong coupling with
the $K\bar K$ channel $1<|g_{f_0K^+K^-}/g_{f_0\pi^+\pi^-}|^2<10$
for the prediction
$|g_{f_0K^+K^-}/g_{f_0\pi^+\pi^-}|^2=(\sqrt{\lambda}-2)^2/4\approx
0.4$. In addition, the mass degeneration $m_{f_0}\approx m_{a_0}$
is coincidental in this case if to treat the $a_0$-meson as the
four-quark state or contradicts the two-quark hypothesis.

The introduction of a gluon component, $gg$, in the $f_0(980)$
meson structure  allows the puzzle of weak coupling with two
gluons and with two photons but the strong coupling with the
$K\bar K$ channel to be resolved easily:
$f_0=gg\sin\alpha\,+$\,$[(1/\sqrt{2})(u\bar u+d\bar
d)\sin\beta+s\bar s\cos\beta]\cos\alpha$,
$\tan\alpha=-O(\alpha_s)(\sqrt{2}\sin\beta +\cos\beta),$ where
$\sin^2\alpha\leq 0.08$ and $\cos^2\beta
> 0.8$. So, the $f_0(980)$ meson is  near to the $s\bar s$-state. It
gives $$0.1 <\frac{B(J/\psi\to f_0(980)\omega)}{B(J/\psi\to
f_0(980)\phi)}= \frac{1}{\lambda}\tan^2\beta<0.54\,.$$ As for the
experimental value, $B(J/\psi\to f_0(980)\omega)/B(J/\psi\to
f_0(980)\phi)=0.44\pm 0.2$, it needs refinement. Remind that in
our opinion the $J/\psi\to f_0(980)\omega$ was not observed!

The scenario with the $f_0(980)$ meson near to the $s\bar s$ state
and with the $a_0(980)$ meson as the two-quark state runs into
following difficulties. 1) It is impossible to explain the $f_0$
and $a_0$-meson mass degeneration in a natural way. 2) It is
predicted
$\Gamma(f_0\to\gamma\gamma)<0.13\cdot\Gamma(a_0\to\gamma\gamma)$,
that means that $f_0(980)$ could not be seen practically in the
$\gamma\gamma$ collision. 3) It is predicted $B(J/\psi\to
a_0(980)\rho)=(3/\lambda\approx 6)\cdot B(J/\psi\to
f_0(980)\phi)$, that has almost no chance from experimental point
view. 4) The $\lambda$ independent prediction $B(J/\psi\to
f_0(980)\phi)/B(J/\psi\to f_2'(1525)\phi)=B(J/\psi\to
a_0(980)\rho)/B(J/\psi\to a_2(1320)\rho)<0.04\pm 0.008$ is
excluded by the central figure in $B(J/\psi\to f_0(980)\phi)
/B(J/\psi\to f_2'(1525)\phi)= 0.4\pm0.23$. But, certainly,
experimental error is too large. Even twofold increase in accuracy
of measurement of could be crucial in the fate of the scenario
under discussion.

 The prospects for the
model of the $f_0(980)$ meson as the almost pure $s\bar s$-state
and the $a_0(980)$-meson as the four-quark state with the
coincidental mass degeneration is rather poor especially as the
OZI-superallowed $\left (N_C\right )^0$ order mechanism $\phi =
s\bar s\to\gamma s\bar s = \gamma f_0(980)$ \footnote{Such a
mechanism is similar to the principal mechanism of the
$\phi\to\gamma\eta '(958)$ decay: $\phi = s\bar s\to\gamma s\bar s
= \gamma\eta'(958)$.} cannot explain the photon spectrum in
$\phi\to\gamma f_0(980)\to\gamma\pi^0\pi^0$ \cite{A2}, which
requires the domination of the $K^+K^-$ intermediate state in the
$\phi\to\gamma f_0(980)$ amplitude: $\phi\to K^+K^-\to\gamma
f_0(980)$! The $\left (N_C\right )^0$ order transition is bound to
have a small weight in the large $N_C$ expansion of the $\phi =
s\bar s\to\gamma f_0(980)$ amplitude, because this term does not
contain the $K^+K^-$ intermediate state, which emerges only in the
next to leading term of the $1/N_C$ order, i.e., in the OZI
forbidden transition \cite{A2}. While the four-quark model with
the symbolic structure $ f_0(980) = (us\bar u\bar s+ds\bar d\bar
s)/\sqrt{2}\cos\theta + ud\bar u\bar d\sin\theta$, similar (but
not identical) the MIT-bag state, reasonably justifies all unusual
features of the $f_0(980)$-meson.

\section{NEW ROUND IN $\gamma\gamma\to\pi^+\pi^-$, THE BELLE DATA \cite{AS2}}

\begin{figure}[h]
\begin{center}
\begin{tabular}{cc}
\includegraphics[width=18pc]{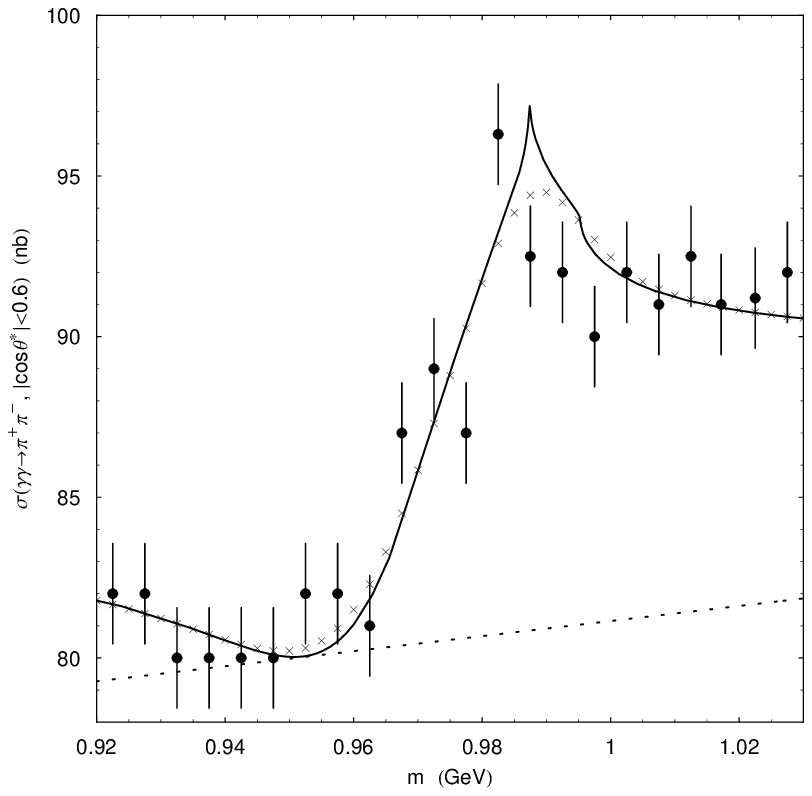}&
\includegraphics[width=18pc]{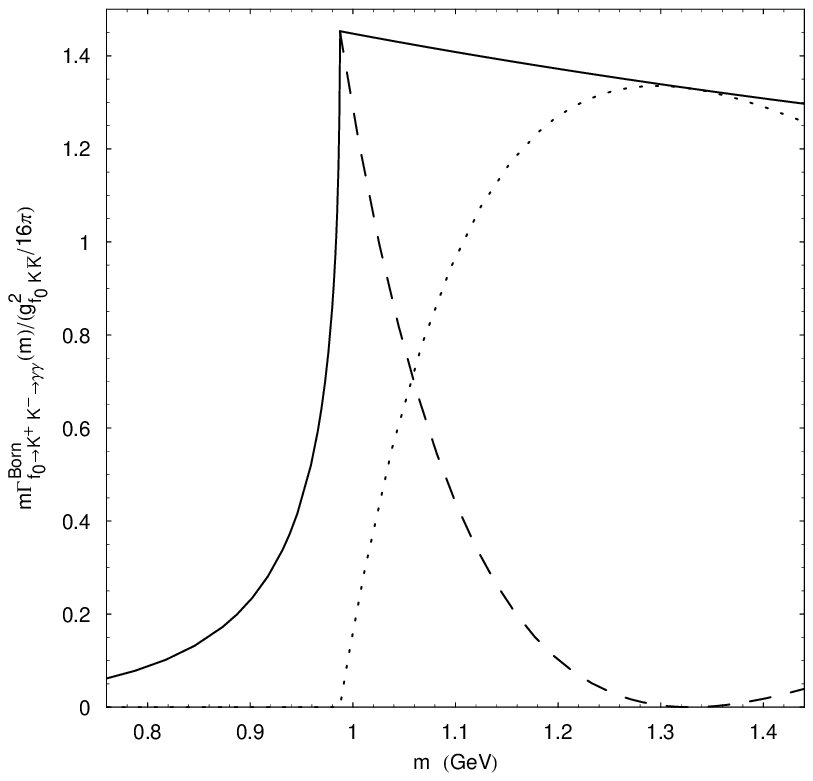}\\
(a)& (b)
\end{tabular}
\end{center}
\caption{\small (a) The fit to the Belle data for  the
$\gamma\gamma\to\pi^+\pi^-$ process . (b) The energy ($m$)
dependence of $f_0\to K^+K^-\to\gamma\gamma$ width. }
 \label{belle}
\end{figure}

Recently, the Belle Collaboration succeeded in observing a clear
manifestation of the $f_0(980)$ resonance in the reaction
$\gamma\gamma\to\pi^+\pi^-$, see Figure \ref{belle} .
 This has been made possible owing to
the huge statistics and good energy resolution.

 Analyzing these
data we shown that the above $K^+K^-$ loop mechanism provides the
absolutely natural and reasonable scale of the $f_0(980)$
resonance manifestation in the $\gamma\gamma\to\pi^+\pi^-$
reaction cross sections as well as in $\gamma\gamma\to\pi^0\pi^0$
\footnote{
 The $a^0_0(980)$ resonance manifestation in $\gamma\gamma\to\pi^0\eta$ is also described by the  $K^+K^-$ loop
 mechanism.}.
  For the
$K^+K^-$ loop mechanism, we obtained the $f_0(980)\to\gamma\gamma$
width averaged by the resonance mass distribution in the $\pi\pi$
channel $\langle\Gamma^{Born}_{f_0\to K^+K^-\to\gamma\gamma}
\rangle_{\pi\pi}\approx0.15$ keV. Furthermore, the $K^+K^-$ loop
mechanism of the $f_0(980)\to\gamma\gamma$ coupling, see Figure
\ref{belle}, is one of the main factors responsible for the
formation of the observed specific, steplike, shape of the
$f_0(980)$ resonance in the $\gamma\gamma\to\pi^+\pi^-$ reaction
cross section.
\section{The  \boldmath{$a_0(980)-f_0(980)$} mixing:  theory and
experiment \cite{ADS}}

 The mixing between the $a_0^0(980)$ and
$f_0(980)$ resonances was discovered theoretically as a threshold
phenomenon in our work in the late 70s. Recently (last decade)
interest in the $a_0^0(980)-f_0(980)$ mixing was renewed, and its
possible manifestations in various reactions are intensively
discussed, because its observation could give an exclusive
information about  the $a_0^0(980)$ and $f_0(980)$ coupling with
the $K\bar K$ channel.

The amplitude of the $a^0_0(980)-f_0(980)$ transition is
determined by the $K^+K^-$ and $K^0\bar K^0$ intermediate states
in the main, $a^0_0(980)\to K^+K^- + K^0\bar K^0\to f_0(980)$,
 \begin{figure}[h]
\begin{center}
\begin{tabular}{cc}
\hspace*{-0.8pc}\includegraphics[width=14pc,height=13.5cm]{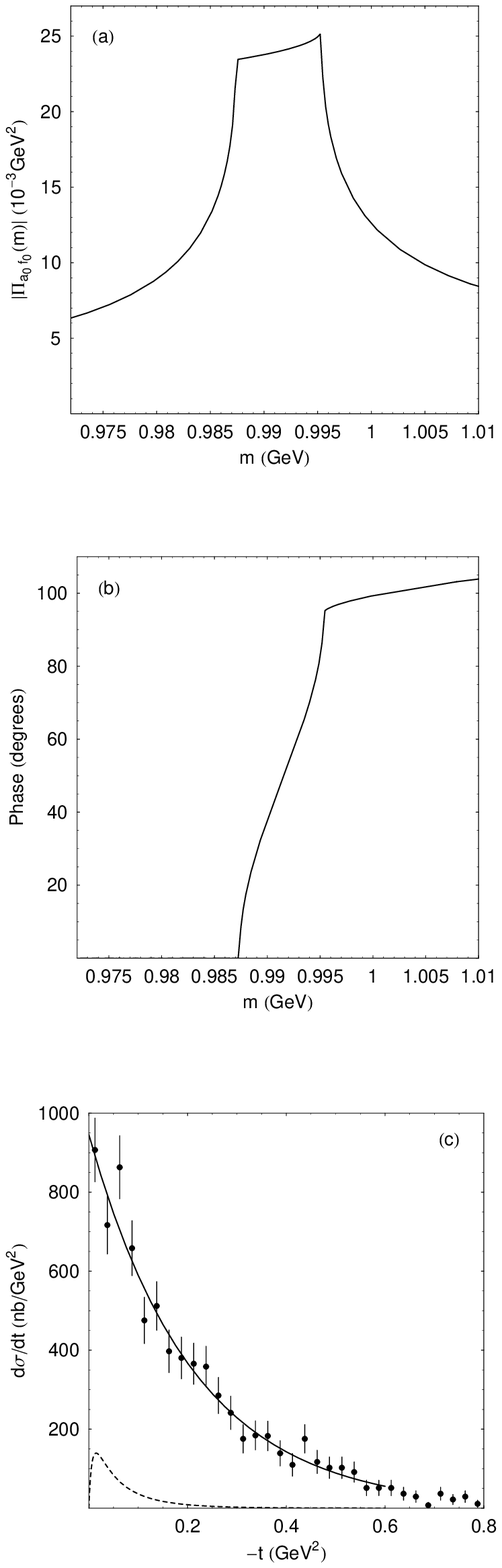}&
\hspace*{-1pc}\includegraphics[width=24pc,height=13.7cm]{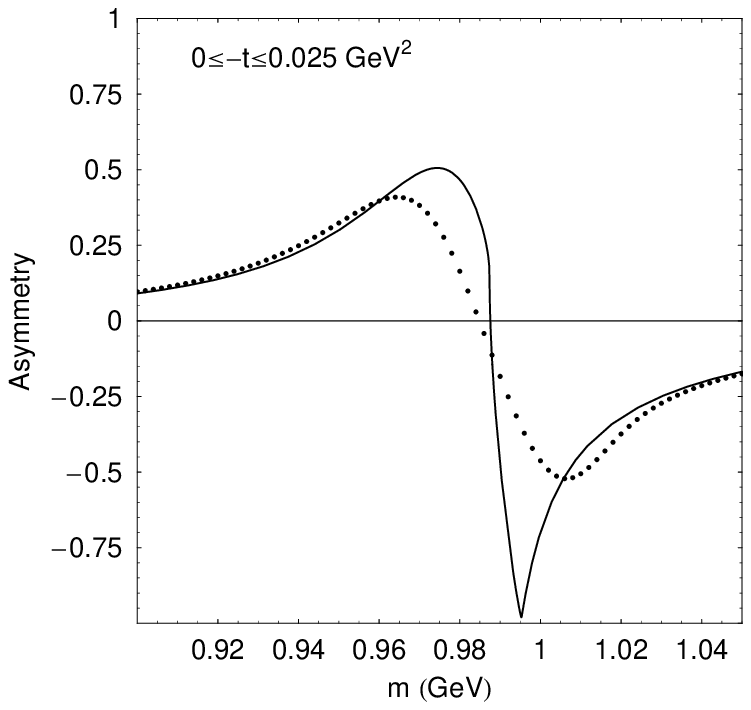}\\
& (d)
\end{tabular}
\end{center}
\caption{\small (a) The modulus of the $a_0-f_0$ transition
amplitude $\Pi_{a_0f_0}(m)$.
 (b) The phase of  the $a_0-f_0$
transition amplitude $\Pi_{a_0f_0}(m)$. (c) The solid curve
describes the nonpolarized $d\sigma/dt(\pi^-p\to a_0n\to\pi^0\eta
n)$ reaction without the $a_0-f_0$ mixing, the dashed curve shows
the $\pi$ exchange contribution due to the $a_0-f_0$ mixing. (d)
The spin asymmetry due to the $a_0-f_0$ mixing, the dotted curve
shows the spin asymmetry smoothed with the Gaussian mass
distribution with the dispersion of 10 MeV.}
 \label{a0f0mixing}
 \end{figure}
\begin{eqnarray}
&&\Pi_{a_0f_0}(m)=\frac{g_{a_0K^+K^-}g_{f_0K^+K^-
}}{16\pi}\Biggl[\,i\,\Bigl(\rho_{K^+K^-}(m)-\rho_{K^0\bar
K^0}(m)\Bigr)-\nonumber\\[9pt]
&&\left.\frac{\rho_{K^+K^-}(m)}{\pi}\ln\frac{1+\rho_{K^+K^-}(m)}
{1-\rho_{K^+K^-}(m)}+\frac{\rho_{K^0 \bar
K^0}(m)}{\pi}\ln\frac{1+\rho_{K^0 \bar K^0}(m)}{1-\rho_{K^0\bar
K^0}(m)}\,\right]\nonumber\\[9pt]
 && \approx
\frac{g_{a_0K^+K^-}g_{f_0K^+K^-
}}{16\pi}\Biggl[\,i\,\Bigl(\rho_{K^+K^-}(m)-\rho_{K^0\bar
K^0}(m)\Bigr)\Biggr], \nonumber
\end{eqnarray}
 where $m\geq2m_{K^0}$, in the
region $0\leq m\leq2m_K$,  $\rho_{K\bar K}(m)=\sqrt{1-4m_K^2/m^2}$
should be replaced by $i|\rho_{K\bar K}(m)|$. In the region
between the $K^+K^-$ and $K^0\bar K^0$ thresholds, which is 8\,MeV
wide,
\begin{eqnarray}
&& |\Pi_{a_0f_0}(m)|\approx
 \frac{|g_{a_0K^+K^-}g_{f_0K^+K^-}|}{16\pi}
\sqrt{\frac{2(m_{K^0}-m_{K^+})}{m_{K^0}}}\nonumber\\[9pt]
 &&\approx
0.127 |g_{a_0K^+K^-}g_{f_0K^+K^-}|/16\pi\gtrsim
0.032\,\mbox{GeV}^2 .\nonumber
\end{eqnarray}

 This contribution dominates  for two reason.\\
  i) It has the
$\sqrt{m_d-m_u}$ order. As for effects of the $m_d-m_u$ order,
they are small. A clear idea of the magnitude of effects of the
$m_d-m_u$ order gives $|\Pi_{a_0f_0}(m)|$ at $m<0.95$ and $m>1.05$
in   Fig. \ref{a0f0mixing} (a). \footnote{ Note that
$|\Pi_{\rho^0\omega}|\approx |\Pi_{\pi^0\eta}|\approx
0.0036\,\mbox{GeV}^2\sim m_d-m_u$.}\\ ii) The strong coupling of
$a_0^0(980)$ and $f_0(980)$ to the $K\bar K$ channels
$|g_{a_0K^+K^-}g_{f_0K^+K^-}|/4\pi\gtrsim 1\,\mbox{GeV}^2$.

 The  "resonancelike" behavior of the $a_0^0(980)-f_0(980)$ mixing
modulus and phase of the amplitude $\Pi_{a_0f_0}(m)$ is clearly
illustrated in  Figs. \ref{a0f0mixing} (a) and (b).

The phase jump suggest the idea to study the $a_0^0(980)-f_0(980)$
mixing in polarization phenomena. If a process amplitude with a
spin configuration is dominated by the $a_0^0(980)-f_0(980)$
mixing then a spin asymmetry of a cross section jumps near the
$K\bar K$ thresholds. An example is $\pi^-p\to \left (a_0^0(980)+
f_0(980)\right )n\to\eta\pi^0\,n$.
\begin{equation}
\hspace*{-7pt}\frac{d^3\sigma}{dtdmd\psi}=
\frac{1}{2\pi}[\,|M_{++}|^2+|M_{+-}|^2 + 2\,\Im
(M_{++}M^*_{+-})\,P\cos\psi\,]\nonumber
\end{equation}

The dimensionless normalized spin asymmetry $$A(t,m)=2\,\Im
(M_{++}M^*_{+-})/[\,|M_{++}|^2+|M_{+-}|^2\,]\,,\quad -1\leq
A(t,m)\leq1\,,$$
 where $M_{+-}$ and $M_{++}$ are the $s$-channel
helicity amplitudes with and without nucleon helicity flip, $\psi$
is the angle between the normal to the reaction plain, formed by
the momenta of the $\pi^-$ and $\eta\pi^0$ system, and the
transverse (to the $\pi^-$ beam axis) polarization of the protons,
$P$ is a degree of this polarization.

As is seen from Fig. \ref{a0f0mixing} (d), the effect of the
$a_0^0(980)-f_0(980)$ mixing in the spin asymmetry is great, and
its observation does not require the high-quality $\pi^0\eta$ mass
resolution, that
 is very important in the problem under discussion.

\section{Conclusion}

Unfortunately, the majority of current investigations of the mass
spectra in  scalar channels does not study particle production
mechanisms. Because of this, such investigations are essentially
preprocessing  experiments, and  the derivable information is very
relative. The progress in understanding the particle production
mechanisms could essentially further our understanding the light
scalar mesons.

Of fundamental importance is production of quark-antiquark pairs
and  hence virtual hadron pairs that forms both resonances and
backgrounds in the light scalar meson region.  Formally it appears
in the necessity of taking into account loop diagrams and
counter-terms essential for correct consideration high
virtualities of intermediate particles in both non-linear and
linear $\sigma$ models. That is why a temptation by a potential
approach is pregnant with artifacts.

Let us show for dessert that we observe the classic two-quark
$\rho$ meson state in its resonance region due to the four-quark
component  of the $\rho$ meson field. Really, the imaginary part
of the $\pi^+\pi^-\to\rho\to\pi^+\pi^-$ amplitude is defined by
the real $\pi^+\pi^-$ intermediate state, i.e., by four-quark
state. But, this amplitude is pure imaginary  at $m=m_\rho$.
Further still the four-quark component of the $\rho$ meson field
dominates at $m_\rho -\Gamma_\rho/2<m<m_\rho -\Gamma_\rho/2$\,.
Let us dwell on this question. The amplitude
\begin{equation}
 A(\pi^+\pi^-\to\rho\to\pi^+\pi^-\,,\,m)=\frac{g^2_{\rho\pi\pi}m^2\rho^2_{\pi\pi}}{D_\rho(m)}\,,\hspace*{8.2cm} (\mbox{A})\nonumber
 \label{rho}
\end{equation}
where $1/D_\rho(m)$ is the $\rho$ meson propagator to the evident
Lorentz structures.
\begin{equation}
\frac{1}{D_\rho(m)}=\frac{1}{m_\rho^2-m^2}+\frac{1}{m_\rho^2-m^2}\Pi_\rho(m)\frac{1}{m_\rho^2-m^2}+
...\,,\hspace*{6.7cm} (\mbox{B})\nonumber
 \label{rhopropagator}
\end{equation}
where $\Pi_\rho(m)$ is the $\pi^+\pi^-$ loop contribution to the
self-energy of the $\rho$ meson ($\rho\to\pi^+\pi^-\to\rho$), but
it is the  four-quark intermediate state contribution. So, the
first term in the right side of Eq. (\mbox{B}) is defined by the
two-quark intermediate state, but the second and other terms mix
two-quark and four-quark degrees of freedom. The infinite series
in  the right side of Eq. (\mbox{B}) lead to
\begin{equation}
\frac{1}{D_\rho(m)}=\frac{1}{m_\rho^2-m^2-\Pi_\rho(m)}\,.\hspace*{10.3cm}
(\mbox{C})\nonumber
\end{equation}
$\Re(\Pi_\rho(m_\rho))=0$, if the $\rho$ meson mass is defined
completely by two constituent quarks. As for
$\imath\Im(\Pi_\rho(m_\rho)=\imath m\Gamma_\rho(m)$, it is defined
completely by the real intermediate $\pi\pi$ state, i.e., by the
four-quark state. Consequently, the four-quark component  of the
$\rho$ meson field dominates in the $\rho$ meson propagator when
$m\approx m_\rho$.

But in case of the $\rho$  coupling with the $\pi\pi$, $K\bar K$
channels,  the $\rho\to\gamma\pi$, $\gamma\eta$ transitions and so
on, the two-quark component of the $\rho$ meson field works.

The $a_0(980)$ and $f_0(980)$ mesons are  a different matter. As
we have seen in Section 4, the $\phi\to\gamma a_0(980)$, $\gamma
f_0(980)$ transitions are defined by the intermediate compact
$K^+K^-$ state, i.e., by the four-quark state.

\begin{center} {\bf\large Acknowledgments}
\end{center}

I thank Organizers of QUARKS-2006 very much for the kind
invitation, the generous hospitality, and the financial support.

This work was supported also in part by the Presidential Grant
NSh-5362.2006.2 for Leading Scientific Schools.

 {\it\large NB}
{\bf\large References} contain mainly author's articles on basis
of which the present talk has been written. Detailed references to
other authors are in these articles.

\end{document}